\documentclass[onecolumn,authoryear]{els-mrw}

\usepackage{amsmath,amssymb,amsfonts,amsthm,makeidx,graphicx}
\usepackage{txfonts}
\usepackage{helvet}
\usepackage{booktabs}
\usepackage{siunitx}

\begin{document}

\chapter{Nuclear fission: from fundamentals to frontiers}\label{chap:fission}

\author[1,2]{M.~R.\ Mumpower}%

\address[1]{\orgname{Obsidian Research}, \orgaddress{Fort Wayne, IN 46835, USA}}
\address[2]{\orgname{University of Notre Dame}, \orgdiv{Department of Physics and Astronomy}, \orgaddress{Notre Dame, IN 46556, USA}}

\maketitle

\begin{abstract}[Abstract]
Nuclear fission, the division of a heavy atomic nucleus into lighter fragments, is a process of fundamental importance to nuclear physics, energy production, and the origin of the elements.
This chapter, part of the Encyclopedia of Nuclear Physics, provides a pedagogical introduction to a broad array of topics in fission physics, progressing from the macroscopic liquid-drop model and microscopic shell corrections through fission dynamics, observables, modern theoretical frameworks, and experimental techniques ranging from traditional double-energy methods to inverse kinematics at radioactive-ion-beam facilities.
The roles of fission in astrophysical nucleosynthesis, nuclear energy, medical isotope production, and national security are discussed, and open questions including the predictive power of microscopic theory and the fission properties of neutron-rich nuclei relevant to heavy element nucleosynthesis are highlighted.
\end{abstract}

\section{Key Points}
\label{sec:keypoints}

\begin{itemize}
    \item \textbf{Fundamental Mechanism and Energetics:} Nuclear fission is the quantum mechanical process where a heavy atomic nucleus divides into lighter fragments (Sec.~\ref{sec:intro}). This splitting is driven by the variation of nuclear binding energy with mass number, resulting in an extraordinary energy release of approximately 200~MeV per fission event (Sec.~\ref{sec:energetics}).
    \item \textbf{Theoretical Foundations:} The phenomenon is governed by the delicate competition between the macroscopic collective behavior of nuclear matter and the microscopic quantum shell structure of individual nucleons (Sec.~\ref{sec:fundamentals}). A liquid drop plus quantum corrections explains critical features like the double-humped fission barrier and the preference for asymmetric mass yields at low incident particle energy among the actinides (Secs.~\ref{sec:shell} and \ref{sec:doublebarrier}).
    \item \textbf{Complex Dynamics:} Fission is an inherently non-equilibrium dynamic process where collective potential energy is continuously converted into internal heat and kinetic energy as the nucleus stretches from a saddle point to scission (Sec.~\ref{sec:dynamics}). Modern theoretical frameworks, such as nuclear density functional theory and Langevin dynamics, are utilized to map and model these complex, multidimensional potential energy surfaces (Secs.~\ref{sec:dft} and \ref{sec:dynamics_theory}).
    \item \textbf{Experimental Observables:} Because the timescales of fission are extraordinarily brief, the dynamics must be inferred from post-scission physical evidence (Secs.~\ref{sec:scission} and \ref{sec:observables}). These essential experimental observables include fragment mass and charge distributions, total kinetic energy, and the characteristic prompt emission of neutrons and gamma rays (Secs.~\ref{sec:yields}, \ref{sec:tke}, and \ref{sec:prompt}).
    \item \textbf{Astrophysical Role:} Fission acts as the upper mass limit for the rapid neutron-capture process ($r$-process) in extreme environments like neutron star mergers, recycling heavy material back into lighter mass nuclei (Sec.~\ref{sec:rprocess}). Furthermore, the radioactive decay and fission of these heavy products provide an efficient heating source that can power the electromagnetic signals of kilonovae (Sec.~\ref{sec:kilonova}).
    \item \textbf{Technological Applications:} Precise knowledge of fission physics and reliable nuclear data are fundamentally required for a wide array of societal applications (Sec.~\ref{sec:applications}). These encompass the design of commercial nuclear power reactors, the production of crucial medical radioisotopes like $^{99}$Mo, the transmutation of nuclear waste, and the advancement of global nonproliferation efforts (Secs.~\ref{sec:energy}, \ref{sec:medical}, \ref{sec:transmutation}, and \ref{sec:security}).
\end{itemize}

\section{Introduction}
\label{sec:intro}

Nuclear fission is the process by which a heavy atomic nucleus divides into lighter nuclei, accompanied by the emission of neutrons, gamma rays, and a tremendous release of energy. 
Although the word ``fission'' was borrowed from biology (where it describes cell division), the underlying physics is entirely quantum mechanical: the outcome is governed by a delicate competition between the macroscopic collective behavior of nuclear matter and the microscopic quantum shell structure of individual protons and neutrons. 

Binary fission, the division into two fragments, is by far the dominant mode. 
Ternary fission, in which a light charged particle such as an alpha particle accompanies the two main fragments, is observed at a rate of a few per thousand fission events. 
True three-body breakup into three fragments of comparable mass is exceedingly rare. 
This chapter focuses primarily on binary fission and its practical significance. 

The process of fission stands at the intersection of fundamental quantum many-body physics, nuclear engineering, and astrophysics, making it one of the most consequential phenomena in all of science.

\subsection{Historical discovery}
\label{sec:history}

The story of fission begins with a chemical puzzle.
In the mid-1930s, several groups sought to create elements heavier than uranium by bombarding it with the recently discovered neutron.
Enrico Fermi's group in Rome reported new radioactivities that were initially attributed to transuranic elements, and Ir\`{e}ne Joliot-Curie's team in Paris obtained similar results that resisted straightforward interpretation.

The resolution came from Berlin.
In December 1938, Hahn and Strassmann performed meticulous radiochemical separations of the products formed when uranium was irradiated with neutrons.
To their astonishment, the products were not radium or transuranic species but \emph{barium}, an element with roughly half the atomic number of uranium \citep{Hahn1939}.
Hahn conveyed the bewildering result to his longtime collaborator Meitner, who had been forced to flee Germany earlier that year.
Over the 1938 Christmas holiday, Meitner and her nephew Frisch recognized that the uranium nucleus had not merely chipped off a fragment but had divided entirely.

Drawing upon the liquid-drop picture of the nucleus advanced by Gamow and Bohr, Meitner and Frisch reasoned that a sufficiently excited uranium nucleus could elongate until electrostatic repulsion overwhelmed the restoring surface tension, splitting the drop in two \citep{Meitner1939}.
Frisch, on the suggestion of biologist William~A.\ Arnold, adopted the term \emph{fission} by analogy with binary cell division.

The implications were recognized with breathtaking speed.
Within weeks, laboratories across Europe and North America confirmed the fission process and demonstrated the emission of secondary neutrons.
Bohr, who had carried the news to the United States in January 1939, and Wheeler developed the comprehensive theoretical framework based on the liquid-drop model \citep{Bohr1939}.
Szilard, Fermi, and others quickly recognized that if each fission event released more than one neutron on average, a self-sustaining chain reaction could be achieved, with consequences for both energy production and weapons.
This realization set in motion the scientific and political events that would culminate in the first controlled chain reaction, demonstrated by Fermi and colleagues at the University of Chicago in December 1942, and ultimately in the Manhattan Project and the dawn of the nuclear age.

\subsection{Energy release}
\label{sec:energetics}

The driving force behind fission is the variation of nuclear binding energy with mass number.
The binding energy per nucleon, $B/A$, rises steeply for light nuclei, peaks near iron ($^{56}$Fe, $B/A\approx 8.8$~MeV), and then gradually declines for heavier species as the growing Coulomb repulsion among protons competes with the attractive strong nuclear force.
For $^{235}$U the value is roughly 7.6~MeV per nucleon, whereas for typical fission fragments with $A\approx 95$--140 it lies near 8.5~MeV per nucleon.

\begin{figure}[htb]
\centering
\includegraphics[width=0.85\textwidth]{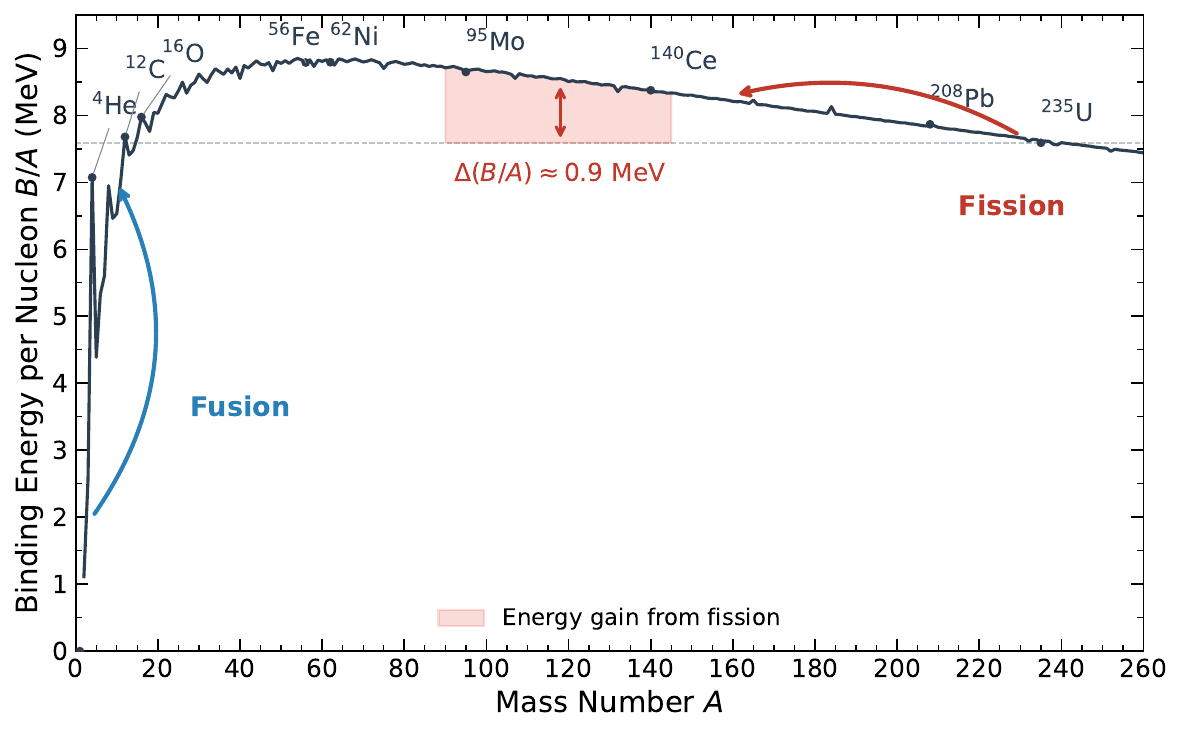}
\caption{Binding energy per nucleon ($B/A$) as a function of mass number
$A$.
The curve peaks near $^{56}$Fe.
For heavy nuclei such as $^{235}$U, the gain of ${\sim}0.9$~MeV per nucleon upon fission to medium-mass fragments drives the enormous energy release.}
\label{fig:be}
\end{figure}

The physical origin of this trend is straightforward.
For light nuclei, adding nucleons increases the number of attractive strong-force bonds faster than it increases Coulomb repulsion, so $B/A$ rises.
Beyond iron, the long-range Coulomb force, which grows as $Z(Z-1)$ and acts between \emph{every} pair of protons rather than just nearest neighbors, gradually overwhelms the short-range strong force, and $B/A$ declines.
This behavior is captured quantitatively by the semi-empirical mass formula (Eq.~\ref{eq:semf}) which is  described in more detail in Section \ref{sec:ldm}. 
The binding energy per nucleus curve is shown graphically in Fig.~\ref{fig:be}.

The curve of $B/A$ versus $A$ immediately reveals two pathways by which nuclei can release energy: \emph{fusion} of light nuclei (climbing the left side of the curve) and \emph{fission} of heavy nuclei (descending the right side).
For fission, when a heavy nucleus splits, each of its roughly 236 nucleons moves from a less tightly bound state to a more tightly bound one.
The difference, approximately 0.9~MeV per nucleon, is liberated as kinetic and radiative energy.
Summed over all nucleons, the total energy release per fission event is $Q \approx 200$~MeV, a figure that exceeds the energy of a typical chemical reaction by roughly six orders of magnitude.
To place this in perspective, the complete fission of one kilogram of $^{235}$U releases approximately $8.2\times 10^{13}$~J, equivalent to the energy content of roughly 2{,}700 metric tons of coal.
This extraordinary energy density is the fundamental reason that nuclear fuel can power a large reactor for over a year from a single loading, and it is also what makes fission relevant to astrophysical phenomena where enormous energy deposition occurs in small volumes of expanding ejecta.

Table~\ref{tab:energy} summarizes the approximate partition of this energy among the various products.
A distinction central to fission physics is that between \emph{prompt} and \emph{delayed} emission.
Prompt products are those released at or very shortly after scission, on timescales of $10^{-20}$--$10^{-14}$~s: the kinetic energy of the two fragments, the neutrons evaporated as the hot fragments cool, and the gamma rays emitted as the excited fragments de-excite.
Delayed products arise from the subsequent radioactive decay of the relatively short-lived fission fragments, occurring over timescales from milliseconds to years; these include beta-decay electrons, antineutrinos, and additional gamma rays.

The lion's share of the total energy appears as kinetic energy of the two charged fragments, which are propelled apart by their mutual Coulomb repulsion.
Prompt neutrons and gamma rays carry away a smaller but experimentally crucial fraction, while the delayed component accounts for roughly 27~MeV released gradually through beta-decay chains.

\begin{table}[htb]
\centering
\caption{Approximate energy partition for thermal-neutron-induced fission
of $^{235}$U \citep{Vandenbosch1973}.}
\label{tab:energy}
\begin{tabular}{lr}
\toprule
Component & Energy (MeV) \\
\midrule
Kinetic energy of fragments & 168 \\
Prompt neutrons             &   5 \\
Prompt gamma rays           &   7 \\
Beta-decay electrons        &   8 \\
Beta-decay antineutrinos    &  12 \\
Delayed gamma rays          &   7 \\
\midrule
\textbf{Total}              & \textbf{$\sim$\,207} \\
\bottomrule
\end{tabular}
\end{table}

\subsection{Why fission matters}
\label{sec:significance}

Fission occupies a distinctive position in contemporary science and technology for at least three reasons.

\textbf{Energy} Among low-carbon electricity sources, nuclear fission is uniquely suited to delivering dispatchable baseload power at the terawatt scale.
Precise knowledge of neutron-induced fission cross sections, prompt neutron spectra, and delayed-neutron fractions underpins the safe design of both current light-water reactors and next-generation systems \citep{GenIV2014}.

\textbf{Nucleosynthesis} The existence of actinides such as uranium and thorium on Earth, in ocean crust samples, and in stellar spectra demonstrates that nature produces nuclei heavy enough for fission to be relevant.
Whether fission actually operates during the rapid neutron-capture process ($r$-process) remains a longstanding open question.
Recent analyses of elemental abundance patterns in metal-poor stars have revealed correlated excesses among certain groups of elements that are consistent with fission fragment deposition, providing the first tantalizing hints that nuclei heavier than uranium were produced and subsequently fissioned in these environments \citep{Roederer2023}.

\textbf{Fundamental physics} Fission presents a grand challenge for nuclear theory: describing the real-time, large-amplitude rearrangement of hundreds of strongly interacting quantum particles.
Progress in density functional theory and high-performance computing has brought microscopic simulations of fission within reach, but predictive capability remains a frontier problem \citep{Schunck2016,Bulgac2016}.

The sections that follow build up the physics of fission from its macroscopic foundations to modern microscopic treatments, survey the key observables and how they are measured, and connect fission to its roles in the cosmos and in technology.
Readers seeking a more detailed treatment are directed to the classic monograph by \citet{Vandenbosch1973}, the theoretical developments in \citet{Krappe2012}, and the modern overview of fission observables, theory, and applications by \citet{Talou2023}; the nuclear many-body methods underpinning microscopic fission theory are covered thoroughly by \citet{Ring1980}; and a modern perspective on the status and challenges of microscopic fission theory is given by \citet{Bender2020}.

\section{Fundamentals of Fission Physics}
\label{sec:fundamentals}

Understanding why and how a nucleus fissions requires bridging two perspectives.
The first treats the nucleus as a charged liquid drop whose bulk properties determine the overall energy landscape.
The second recognizes that protons and neutrons occupy discrete quantum states whose arrangement modulates that landscape in decisive ways.

\subsection{The liquid-drop model}
\label{sec:ldm}

Shortly after the discovery of fission, Bohr and Wheeler cast the problem in terms of the liquid-drop model (LDM) \citep{Bohr1939}.
Nuclear matter is treated as an incompressible, uniformly charged fluid.
The total binding energy of a spherical nucleus with $Z$ protons and $N=A-Z$ neutrons is expressed through the semi-empirical mass formula (also known as the Bethe--Weizs\"{a}cker formula):
\begin{equation}
  E_B(Z,A) = a_V A - a_S A^{2/3} - a_C \frac{Z(Z-1)}{A^{1/3}}
- a_{\rm sym}\frac{(A-2Z)^2}{A} + \delta(A,Z)\,,
  \label{eq:semf}
\end{equation}
where $a_V\approx 15.5$~MeV (volume), $a_S\approx 16.8$~MeV (surface), $a_C\approx 0.72$~MeV (Coulomb), and $a_{\rm sym}\approx 23$~MeV (symmetry) are empirically determined coefficients. 
The first (volume, $E_V$) term on the right hand side reflects the saturation of the nuclear force (each nucleon interacts only with its nearest neighbors); the second (surface, $E_S$) term corrects for nucleons near the surface that have fewer neighbors; the third (Coulomb, $E_C$) term accounts for the electrostatic repulsion among all proton pairs; the fourth (symmetry, $E_\text{sym}$) term penalizes imbalances between proton and neutron numbers \citep{Weizsacker1935}; the final term, $\delta(A,Z)$,  is a pairing correction that favors even-Z, even-N nuclei \citep{Krane1988}. 

For the purpose of understanding fission, the competition between the surface energy $E_S$ (which favors a compact sphere) and the Coulomb energy $E_C$ (which favors expansion and ultimately rupture) is paramount.

Because nuclear matter is treated as an incompressible fluid, a deformed nucleus has the same volume as a sphere of radius $R_0 = r_0\,A^{1/3}$ (with $r_0\approx 1.2$~fm). 
This relation encodes a fundamental observation. 
The parameter $r_0$ represents the approximate effective radius of a single nucleon, acting as the proportionality constant that scales the overall nuclear radius $R_0$ against the cube root of the mass number $A$, which demonstrates that nuclear matter has a roughly constant density. 
The nuclear surface can be described by an angular modulation of that radius,
\begin{equation}
  R(\theta,\phi) = R_0\biggl[1 + \sum_{\lambda=1}^{\infty}\sum_{\mu=-\lambda}^{\lambda}
  \alpha_{\lambda\mu}\,Y_{\lambda\mu}(\theta,\phi)\biggr]\,,
  \label{eq:shape_general}
\end{equation}
where the $Y_{\lambda\mu}$ are spherical harmonics and the coefficients $\alpha_{\lambda\mu}$ specify the magnitude of each multipole distortion.
The $\lambda=0$ term is fixed by volume conservation, and $\lambda=1$ corresponds to a translation of the center of mass rather than a shape change.
The leading true deformation is the \emph{quadrupole} ($\lambda=2$) mode: positive $\alpha_{20}$ stretches the nucleus into a prolate (cigar-like) shape, while negative $\alpha_{20}$ compresses it into an oblate (disk-like) shape.
Higher multipoles ($\lambda=3$ for octupole or pear-shaped distortions, $\lambda=4$ for hexadecapole, etc.)\ become important at larger deformations but can be neglected for the small perturbations considered here.

For an axially symmetric quadrupole deformation it is conventional to write $\alpha_{20}\equiv\beta_2$ (the quadrupole deformation parameter), reducing Eq.~\eqref{eq:shape_general} to
\begin{equation}
  R(\theta) = R_0\bigl[1 + \beta_2\,Y_{20}(\theta)\bigr]\,.
  \label{eq:surface}
\end{equation}
An equivalent form uses the Legendre polynomial $P_2(\cos\theta) = \sqrt{4\pi/5}\;Y_{20}$, absorbing the numerical prefactor into the deformation amplitude; in the discussion that follows we work with $\alpha_2$ defined through $R(\theta)=R_0[1+\alpha_2\,P_2(\cos\theta)]$ so that the algebra is transparent.
The two conventions are related by $\alpha_2 = \sqrt{5/4\pi}\;\beta_2$.
Both appear widely in the literature; $\beta_2$ is the more common notation in modern nuclear structure work and will be used in subsequent sections of this chapter.

When the nucleus deforms, the surface area increases, raising $E_S$; at the same time the average separation between protons grows, lowering $E_C$.
To quantify this competition for small quadrupole deformations, one expands the surface area and electrostatic self-energy to second order in $\alpha_2$, giving \citep{Bohr1939}
\begin{equation}
  E_S = E_S^{(0)}\Bigl[1 + \tfrac{2}{5}\,\alpha_2^2 + \cdots\Bigr]\,,
  \qquad
  E_C = E_C^{(0)}\Bigl[1 - \tfrac{1}{5}\,\alpha_2^2 + \cdots\Bigr]\,,
  \label{eq:ESEC}
\end{equation}
where $E_S^{(0)} = a_S\,A^{2/3}$ and $E_C^{(0)} = a_C\,Z(Z-1)/A^{1/3}$ are the spherical values.
The factor $+2/5$ reflects the growth of the surface area with elongation, while $-1/5$ reflects the reduction in Coulomb energy as the charge distribution spreads.
The net change in macroscopic energy is therefore
\begin{equation}
  \Delta E_{\rm macro}(\alpha_2)
  = \tfrac{1}{5}\bigl(2\,E_S^{(0)} - E_C^{(0)}\bigr)\,\alpha_2^2\,.
  \label{eq:deformation}
\end{equation}

Equation~\eqref{eq:deformation} shows that the nucleus is stable against small deformations when $2E_S^{(0)} > E_C^{(0)}$ and unstable when the inequality reverses.
Bohr and Wheeler codified this balance through the fissility parameter
\begin{equation}
  x = \frac{E_C^{(0)}}{2\,E_S^{(0)}}
\approx \frac{Z^2/A}{(Z^2/A)_{\rm crit}}\,,
  \label{eq:fissility}
\end{equation}
where the superscript $(0)$ denotes the spherical configuration and the critical value $(Z^2/A)_{\rm crit}\approx 47$--$51$ depending on the parameterization of the surface and Coulomb energies.
When $x<1$, the spherical shape is a local energy minimum and a finite energy barrier must be surmounted for fission to proceed.
When $x\geq 1$, surface tension can no longer contain the Coulomb pressure and the nucleus disintegrates without any barrier.
All naturally occurring nuclei have $x<1$; for instance, $^{235}$U gives $x\approx 0.7$--$0.8$.

The fissility parameter provides a useful organizing principle.
For light nuclei ($x\ll 1$), fission barriers are enormous and the process is energetically irrelevant.
As $Z$ increases through the rare-earth region ($x\sim 0.5$), barriers remain substantial ($>20$~MeV) and fission is observed only at the highest excitation energies.
In the actinide region ($x\sim 0.7$--0.8), barriers have dropped to ${\sim}5$--6~MeV and compete with neutron separation energies, making neutron-induced fission a dominant reaction channel.
Beyond the actinides, extrapolation toward $x=1$ suggests that the periodic table must eventually end when the barrier vanishes entirely. The pure LDM places this limit near element $Z\approx 104$ \citep{Bohr1939,Krane1988}.
As will be shown, quantum shell effects can extend this limit substantially (Sec.~\ref{sec:outlook}).

The pure LDM predicts a single, smooth fission barrier whose height diminishes as $x$ approaches unity.
For a typical actinide the predicted barrier height is roughly 6~MeV, comparable to the neutron separation energy $S_n$.
The interplay between barrier height and separation energy determines which isotopes can be fissioned by thermal neutrons, a point returned to in Sec.~\ref{sec:spontaneous}.

\subsection{Shell corrections and the Strutinsky method}
\label{sec:shell}

Despite its successes, the LDM fails in two striking ways.
First, it predicts that the energetically most favorable split is into two equal-mass fragments (symmetric fission), yet low-energy fission of actinides is overwhelmingly asymmetric. For example, $^{235}$U preferentially divides into a heavier fragment near $A\approx 140$ and a lighter one near $A\approx 95$.
Second, the LDM cannot account for the anomalously long spontaneous-fission half-lives of certain isotopes or the existence of shape isomers (see Sec.~\ref{sec:doublebarrier}).

The resolution lies in quantum shell structure.
In a manner analogous to the enhanced stability of closed electron shells in atoms, nuclei acquire additional binding energy when their proton or neutron shells are fully occupied.
Nucleons move in a mean-field potential generated by all the other nucleons, and the resulting single-particle energy spectrum exhibits gaps at certain occupation numbers, the ``magic numbers'' 2, 8, 20, 28, 50, 82, and 126. 
Nuclei with magic proton or neutron numbers (``singly magic'') or both (``doubly magic,'' such as $^{132}$Sn with $Z=50$ and $N=82$) exhibit enhanced binding energy, higher first excited-state energies, and smaller radii compared to their neighbors.

The relevance to fission is that these single-particle energy levels shift as the nuclear shape changes.
A configuration that is magic at spherical symmetry may lose that character upon deformation, while new ``deformed shell gaps'' can open at other shapes.
The density of single-particle levels near the Fermi surface (the highest occupied energy level) fluctuates as a function of shape, producing oscillations in the total nuclear energy that are superimposed on the smooth liquid-drop trend.

In the late 1960s, Strutinsky devised an elegant scheme to marry the macroscopic and microscopic pictures \citep{Strutinsky1967,Strutinsky1968}.
He proposed writing the total nuclear energy as
\begin{equation}
  E_{\rm total}(\boldsymbol{q})
= E_{\rm LDM}(\boldsymbol{q}) + \delta E_{\rm shell}(\boldsymbol{q}) + \delta E_{\rm pair}(\boldsymbol{q})\,,
  \label{eq:strutinsky}
\end{equation}
where $\boldsymbol{q}$ denotes the set of deformation coordinates.
The shell correction $\delta E_{\rm shell}$ is obtained by subtracting a smoothed single-particle level density from the actual, fluctuating level density; the pairing correction $\delta E_{\rm pair}$ accounts for the superfluid correlations among nucleon pairs.
This \emph{macroscopic--microscopic} (mac-mic) method preserves the correct bulk properties of nuclear matter while capturing the quantum oscillations that govern the fine structure of the fission path.

\subsection{The double-humped fission barrier}
\label{sec:doublebarrier}

\begin{figure}[htb]
\centering
\includegraphics[width=0.85\textwidth]{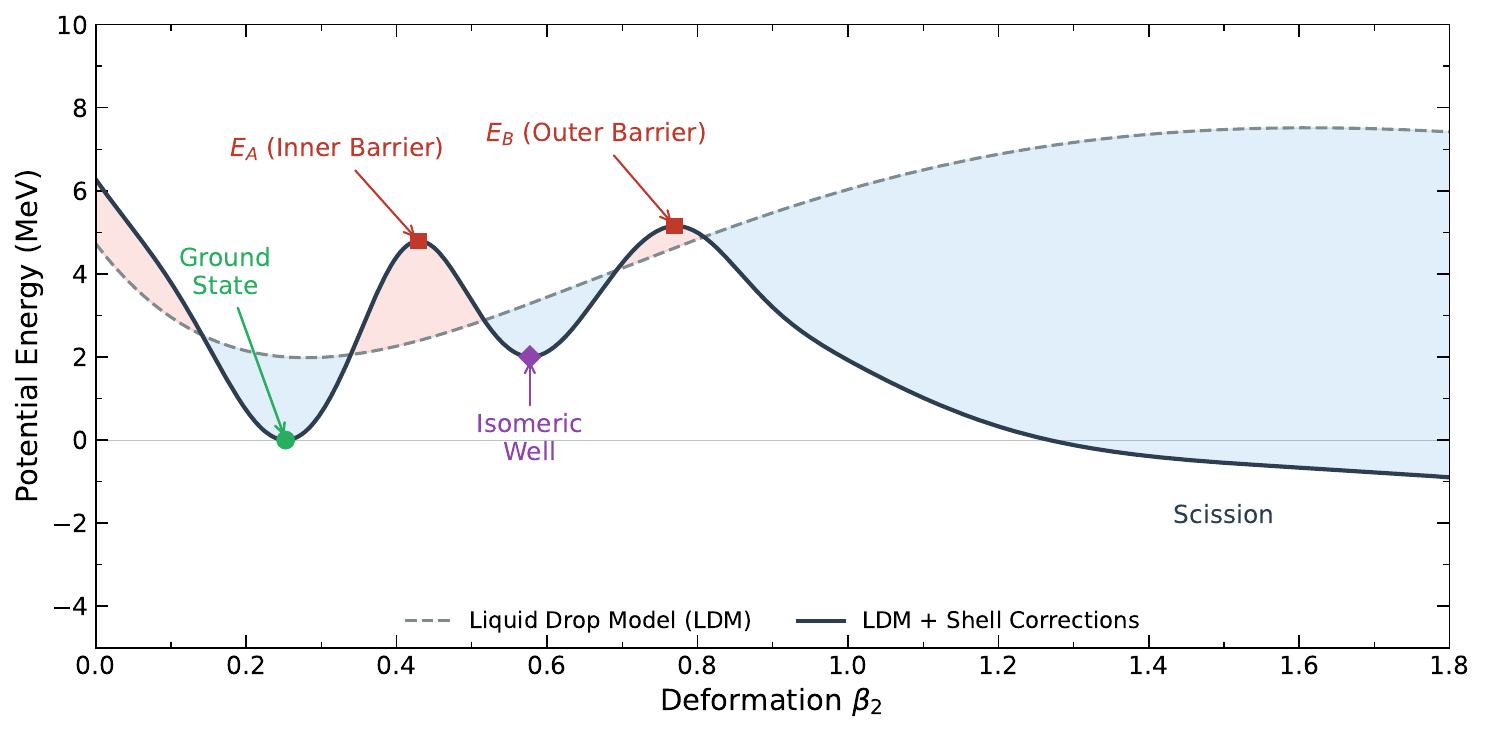}
\caption{Schematic of the double-humped fission barrier for a typical
actinide nucleus.
The dashed line shows the smooth liquid-drop model (LDM) prediction; the solid line includes shell corrections, which create the inner barrier ($E_A$), the isomeric second minimum, and the outer barrier ($E_B$).
Blue shading indicates regions where shell effects lower the energy (stabilizing), while red shading indicates regions where they raise it (destabilizing).
Beyond the outer barrier the energy descends toward scission, the point at which the two nascent fragments separate and the nuclear force no longer bridges them.
This one-dimensional cross-section is shown for illustration; modern calculations employ two to five collective coordinates (elongation, mass asymmetry, neck thickness, fragment deformation) to capture the full multidimensional topography of the fission path (see Sec.~\ref{sec:pes}).}
\label{fig:barrier}
\end{figure}

Applying the Strutinsky prescription to actinide nuclei reveals a dramatic departure from the smooth LDM barrier.
Shell corrections sculpt the barrier into a \emph{double-humped} profile with several distinctive features (see Fig.~\ref{fig:barrier}):

\begin{enumerate}
\item \textbf{Ground-state minimum}  The deformed ground state of the
actinides (near $\beta_2\approx 0.25$) owes its stability to a large negative shell correction that deepens the energy well relative to the LDM prediction.

\item \textbf{Inner barrier ($E_A$)}  Stretching the nucleus beyond the
ground state leads to a region of low shell stabilization, creating a first barrier peak.

\item \textbf{Second minimum (isomeric well)}  At a 2:1 axis ratio
($\beta_2\approx 0.6$), a new set of deformed shell gaps produces a strong negative shell correction, carving out a secondary potential well typically 2--3~MeV above the ground state.

\item \textbf{Outer barrier ($E_B$)}  Beyond the isomeric well a
second barrier peak must be traversed before the energy drops irreversibly toward scission.
\end{enumerate}

This structure immediately explains the existence of \emph{fission isomers} (or shape isomers): nuclei trapped in the secondary minimum are metastable, highly deformed states that can decay by spontaneous fission through the comparatively thin outer barrier, giving rise to anomalously short fission half-lives, first observed experimentally by \citet{Polikanov1968,Vandenbosch1977} and systematized in the comprehensive review by \citet{Bjornholm1980}.
The double barrier also accounts for the vibrational resonance structure observed in sub-barrier neutron-induced fission cross sections, where quantum tunneling through the barrier proceeds via intermediate states in the second well.

\subsection{Potential energy surfaces}
\label{sec:pes}

\begin{figure}[htb]
\centering
\includegraphics[width=0.85\textwidth]{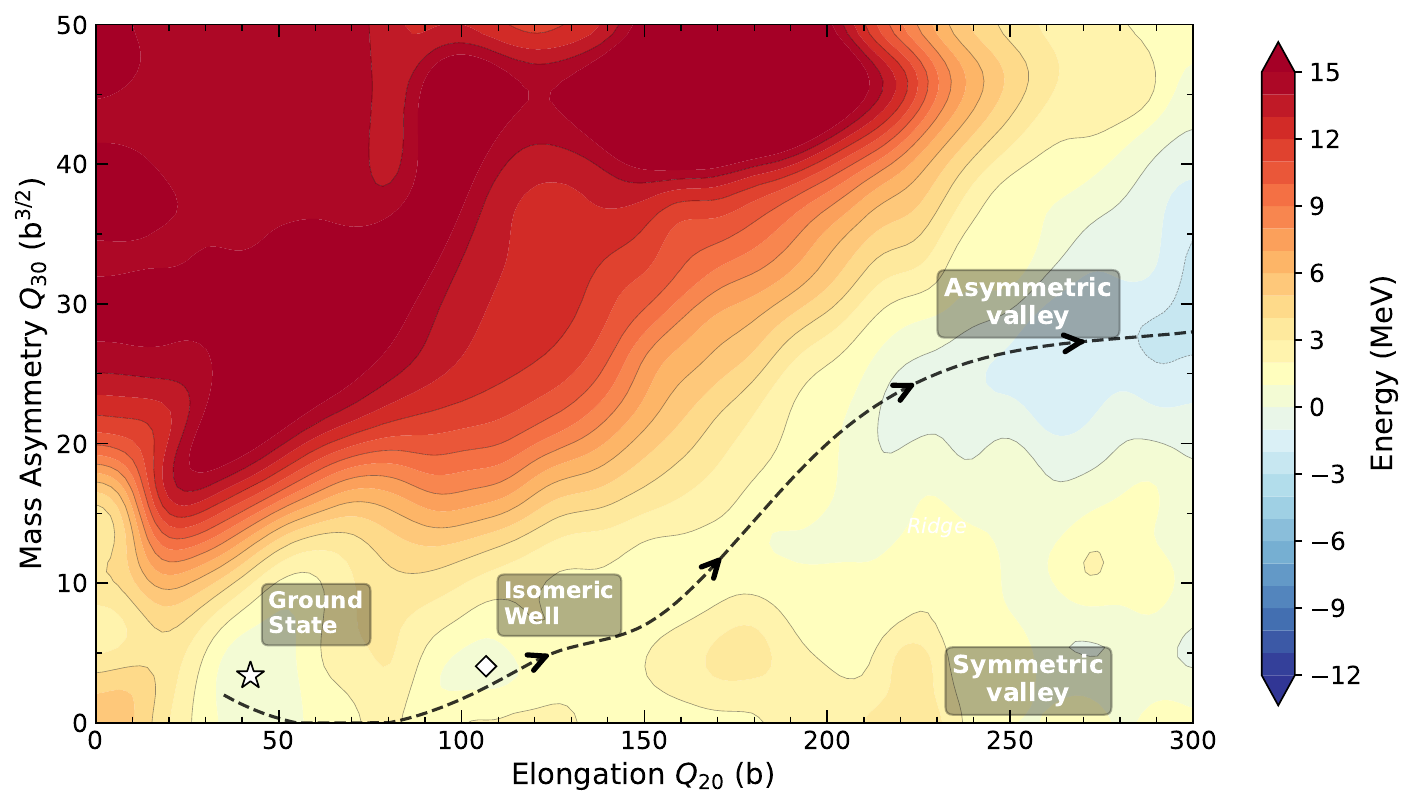}
\caption{Schematic two-dimensional potential energy surface for a representative
actinide nucleus as a function of elongation ($Q_{20}$) and mass asymmetry ($Q_{30}$).
The ground state, isomeric well, and the asymmetric and symmetric fission valleys are indicated.
The dashed line traces a representative fission path from the ground state, over the inner and outer barriers, and into the asymmetric valley, which lies lower in energy and explains the dominance of asymmetric mass splits at low excitation energy.
This figure is illustrative; actual PES topographies are obtained from theoretical calculations (see Sec.~\ref{sec:dft}).}
\label{fig:pes}
\end{figure}

A single deformation coordinate is insufficient to describe the full complexity of fission.
Modern calculations map the nuclear energy across a multidimensional deformation space, typically three to five collective coordinates characterizing elongation, mass asymmetry (the relative sizes of the two nascent fragments), neck thickness, and fragment deformation.
The resulting \emph{potential energy surface} (PES) is analogous to a topographic map of a mountain landscape (Fig.~\ref{fig:pes}).

The ground state resides in a valley; fission requires traversal of mountain passes (saddle points) corresponding to the inner and outer barriers.
Beyond the outer saddle, the topography typically bifurcates into distinct \emph{fission valleys}: a symmetric valley at zero mass asymmetry and one or more asymmetric valleys at nonzero asymmetry.
For most actinides the asymmetric valley lies several MeV lower in energy, explaining the dominance of asymmetric mass splits in low-energy fission of uranium and plutonium isotopes \citep{Moller2001}.
A prototypical example is the PES of $^{236}$U, the compound nucleus formed in $^{235}$U~+~$n$ (Fig.~\ref{fig:pes}).

Different fissioning nuclei present qualitatively different PES topographies, with the relative depths of the symmetric and asymmetric valleys varying across the chart of nuclides (specific examples are discussed in Sec.~\ref{sec:yields}).
The central insight from decades of PES calculations is that the outcome of fission is a direct manifestation of nuclear shell structure, not just in the fissioning parent, but in the fragments that will emerge.
This represents one of the deepest connections between nuclear structure and nuclear reactions.

The PES encodes the \emph{statics} of fission: the energetic landscape the nucleus must navigate.
How the nucleus actually traverses this landscape is the subject of fission dynamics.

\section{Fission Dynamics}
\label{sec:dynamics}

If the potential energy surface provides the map, fission dynamics describes the journey.
Fission is inherently a non-equilibrium process: collective potential energy is continuously converted into intrinsic excitation (heat) and kinetic energy as the nucleus evolves from a compact shape through scission to two separated fragments.

\subsection{The statistical picture: fission as a decay channel}
\label{sec:statistical}

At the most basic level, fission of an excited compound nucleus can be treated as one among several competing decay channels.
When a nucleus absorbs a neutron and forms a compound system at excitation energy $E^*$, the compound nucleus ``forgets'' how it was formed (Bohr's independence hypothesis) and decays according to statistical widths.
The fission width is given approximately by the Bohr--Wheeler transition-state formula \citep{Bohr1939}:
\begin{equation}
\Gamma_f(E^*) = \frac{1}{2\pi\,\rho_{\rm gs}(E^*)} \int_0^{E^*-B_f} \rho_{\rm sad}(E^* - B_f - \varepsilon)\, d\varepsilon\,,
  \label{eq:bohrwheeler}
\end{equation}
where $\rho_{\rm gs}$ is the level density at the ground-state deformation, $\rho_{\rm sad}$ is the level density at the saddle point, $B_f$ is the fission barrier height, and natural units ($\hbar=1$) are used.
Physically, the formula counts the number of quantum states available at the saddle-point configuration: the more states accessible above the barrier, the wider the fission channel and the faster the nucleus fissions.
The competing neutron-emission width involves the level density of the daughter nucleus and the neutron transmission coefficients.

The ratio $\Gamma_f/\Gamma_n$ (fission to neutron emission) determines the fission probability at each excitation energy.
Whether this ratio is large or small relative to neutron emission depends on the relationship between $B_f$ and $S_n$, as discussed in the next subsection.

\subsection{Modes of fission initiation}
\label{sec:spontaneous}

Before discussing the descent to scission, it is important to distinguish the principal mechanisms by which fission can be initiated.

\textbf{Spontaneous fission} occurs when a nucleus in its ground state tunnels quantum mechanically through the fission barrier.
The process is governed by the barrier height, width, and the nuclear inertia along the fission path.
Spontaneous-fission half-lives vary enormously: from ${\sim}2\times 10^{17}$~years for $^{238}$U to fractions of a microsecond for the heaviest transfermium nuclei.
The extreme sensitivity to barrier parameters makes spontaneous fission an exacting test of potential energy surface calculations.

\textbf{Induced fission} is triggered by supplying excitation energy to the nucleus, most commonly through neutron capture but also via photon absorption (photofission), charged-particle reactions, or Coulomb excitation.
Whether a given neutron energy is sufficient to induce fission depends on the relationship between the fission barrier height $B_f$ and the neutron separation energy $S_n$ of the resulting compound nucleus.
For $^{235}$U, absorption of a low-energy neutron forms the compound nucleus $^{236}$U$^*$ with an excitation energy of $S_n \approx 6.5$~MeV, which exceeds $B_f \approx 5.8$~MeV; hence $^{235}$U is \emph{fissile} with thermal neutrons.
By contrast, $^{238}$U has $S_n \approx 4.8$~MeV for thermal neutron capture (forming $^{239}$U$^*$), which falls below its barrier of ${\sim}5.9$~MeV, making it \emph{fissionable} only with fast neutrons above roughly 1~MeV.
This distinction between fissile and fissionable isotopes is central to reactor physics.

The neutron-induced fission cross section $\sigma_f(E_n)$ encodes this competition quantitatively as a function of incident neutron energy.
For a fissile nucleus like $^{235}$U, $\sigma_f$ is large (${\sim}580$~b) at thermal energies, exhibits a dense forest of individual compound-nuclear resonances in the eV--keV range, and then settles to a smooth plateau of ${\sim}1$--2~b above a few MeV.
For a fissionable but non-fissile nucleus like $^{238}$U, $\sigma_f$ is negligible below the effective threshold near 1~MeV, then rises steeply and levels off at a comparable plateau.
Accurate knowledge of $\sigma_f(E_n)$ across this full energy range is essential: the thermal and resonance regions govern reactor criticality calculations, while the high-energy regions are relevant to fast reactor design and national security. 

\textbf{Beta-delayed fission} provides a third pathway.
A neutron-rich or neutron-deficient nucleus that is itself stable against fission may beta-decay to excited states in the daughter nucleus.
If the excitation energy of the populated level exceeds the daughter's fission barrier, the daughter can fission before de-exciting by neutron or gamma-ray emission.
Beta-delayed fission is rare under ordinary conditions because the $Q_\beta$ values of known nuclei seldom reach the fission barrier of the daughter.
However, for the very neutron-rich nuclei encountered in the $r$-process (Sec.~\ref{sec:rprocess}), $Q_\beta$ values grow to tens of MeV while fission barriers shrink, making beta-delayed fission a potentially significant decay channel \citep{Thielemann1983}. 
Presently, the only experimental confirmation of beta-delayed fission resides for neutron-deficient nuclei \citep{Andreyev2013}.

\subsection{From saddle to scission}
\label{sec:saddle2scission}

Once a nucleus, whether by tunneling or by excitation above the barrier, crosses the outer most saddle point, it enters a steep descent on the PES.
During this phase the neck connecting the two lobes thins progressively while the nascent fragments draw apart.
The collective kinetic energy gained on the downslope is partially dissipated into single-particle degrees of freedom, heating the nucleus internally.

The dynamics of the descent can be understood by analogy.
Imagine a ball rolling down a rugged mountain valley filled with a viscous fluid.
The ball (representing the collective nuclear shape) is driven downhill by gravity (the gradient of the PES) but retarded by friction (nuclear dissipation) and buffeted by random thermal kicks (fluctuations).
This picture forms the physical basis of the Langevin approach discussed in Sec.~\ref{sec:dynamics_theory}.

Two dissipation mechanisms are commonly discussed.
\emph{Two-body dissipation} arises from nucleon--nucleon collisions, analogous to classical viscosity in a fluid.
\emph{One-body dissipation} (or wall-and-window friction) results from nucleons scattering off the time-dependent mean-field potential as the nuclear shape evolves; it is generally regarded as the dominant mechanism in large-amplitude collective motion which occurs during nuclear fission \citep{Blocki1978}.

The distinction matters quantitatively.
One-body dissipation predicts stronger friction and thus more internal heating of the fragments during the descent, leading to lower collective kinetic energies at scission.
Two-body dissipation, being weaker, allows the nucleus to arrive at scission with more collective kinetic energy but lower fragment temperatures.
The measured partition of the total energy release between fragment kinetic energy and excitation energy therefore provides constraints on the dissipation mechanism.

An important consequence of dissipation is the concept of \emph{nuclear viscosity}.
Early estimates of the saddle-to-scission time, based on undamped motion on the PES, predicted transit times as short as $10^{-21}$~s.
With realistic one-body dissipation, the transit time increases by roughly an order of magnitude, and the nucleus can equilibrate its shape degrees of freedom during the descent.
This ``overdamped'' regime, where the nucleus creeps rather than slides down the PES, is the physical basis for the success of Langevin approaches that treat the collective motion as diffusive rather than ballistic.
Experimentally, the degree of damping has been probed through pre-scission neutron multiplicities and through the suppression of fission at high angular momentum, both of which are sensitive to the time spent on the descent.

\subsection{Scission and timescales}
\label{sec:scission}

Scission marks the irreversible rupture of the neck, the moment at which the strong nuclear force ceases to bridge the two fragments.
The concept is intuitive but surprisingly subtle: in quantum mechanics the transition from one connected body to two separated ones is not a sharp classical event.
Recent microscopic calculations have shown that scission involves the rapid tunneling of nucleon wave functions from the neck region into the fragments, occurring over ${\sim}10^{-22}$~s \citep{Bulgac2016}.

As the neck ruptures, the nascent fragments are typically highly deformed, stretched by their mutual Coulomb attraction during the descent, and internally excited from dissipative heating.
Once fully separated, they reach roughly 90\% of their asymptotic kinetic energy within ${\sim}10^{-20}$~s.

The timescales governing fission span several orders of magnitude:
\begin{itemize}
\item \textbf{Compound nucleus lifetime} (${\sim}10^{-19}$--$10^{-15}$~s):
the time a thermally excited nucleus spends before ``deciding'' to fission rather than emit a neutron or gamma ray.
This is governed by the ratio of the fission width $\Gamma_f$ to the total decay width.
\item \textbf{Transient time} (${\sim}10^{-21}$--$10^{-20}$~s): the
interval required to build up probability flux across the barrier at a given excitation energy.
\item \textbf{Saddle-to-scission time} (${\sim}10^{-20}$~s): the
duration of the descent from the barrier to the point of neck rupture.
\item \textbf{Post-scission acceleration} (${\sim}10^{-20}$~s):
Coulomb-driven separation to the asymptotic fragment velocities.
\end{itemize}
These extraordinarily brief timescales place direct observation of the scission configuration beyond current experimental capabilities.
Instead, the dynamics must be inferred from the properties of the products that emerge: the fission observables.

\section{Fission Observables}
\label{sec:observables}

The observables of fission, including fragment masses, charges, kinetic energies, and prompt radiation, serve as forensic evidence from which the incredibly fast dynamics of the process can be reconstructed.

\subsection{Fragment mass and charge distributions}
\label{sec:yields}

\begin{figure}[htb]
\centering
\includegraphics[width=0.85\textwidth]{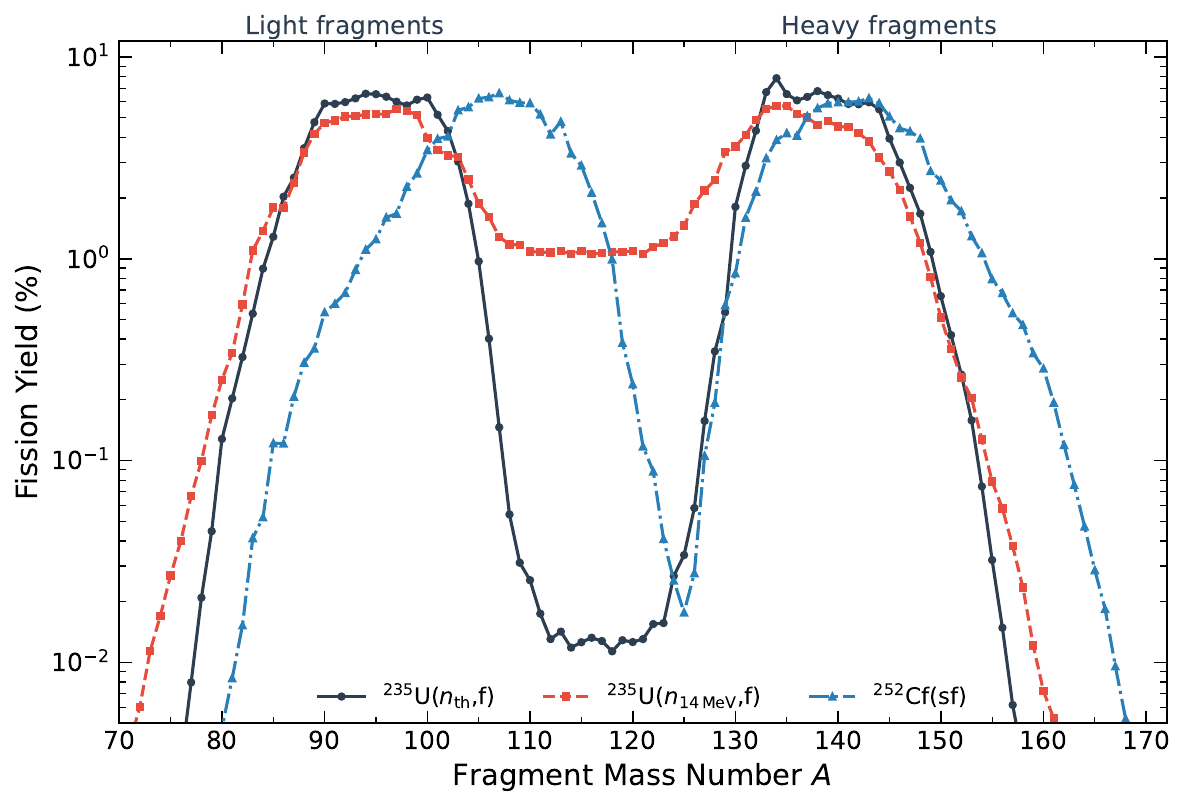}
\caption{Fission fragment mass yield distributions from the ENDF/B-VIII.0 evaluation.
The thermal-neutron-induced fission of $^{235}$U shows the classic asymmetric double-humped pattern, which evolves toward symmetry at higher excitation energies (14~MeV neutrons).
The spontaneous fission of $^{252}$Cf displays a similar asymmetric distribution shifted to higher fragment masses owing to the larger mass of the fissioning system.}
\label{fig:yields}
\end{figure}

When an actinide such as $^{235}$U undergoes low-energy fission, the resulting mass distribution is not a single peak centered at half the parent mass.
Instead, it exhibits a characteristic double-humped shape (Fig.~\ref{fig:yields}): the yield is concentrated around a heavy fragment with $A\approx 140$ and a complementary light fragment with $A\approx 95$, with a deep minimum at symmetry ($A\approx 118$).

The position of the heavy peak is remarkably stable across the actinide series, from uranium through californium.
This robustness is attributed to the strong spherical shell closures at proton number $Z=50$ and neutron number $N=82$, which endow the heavy fragment with extra binding energy and pull the mass split toward that favorable configuration \citep{Wilkins1976}.
The light fragment peak shifts to accommodate the changing total mass.
A useful phenomenological framework for describing this structure is the multi-modal fission model of Brosa and collaborators \citep{Brosa1990}, which decomposes the yield into several independent ``channels'': the \emph{Standard~I} and \emph{Standard~II} modes, each associated with a distinct scission configuration and shell-stabilized fragment pair, and a broad \emph{Superlong} mode corresponding to highly elongated, near-symmetric splits.
The relative weights of these modes vary with excitation energy and fissioning system, providing a compact parameterization of the yield data.

At higher excitation energies the shell effects are progressively washed out by thermal fluctuations.
The valley between the asymmetric peaks fills in, and the yield distribution evolves toward a single symmetric peak, a manifestation of the underlying liquid-drop preference for equal division \citep{Andreyev2018,Schmidt2018}.
Conversely, very heavy actinides such as $^{258}$Fm exhibit symmetric fission even at low energy, because both fragments can simultaneously achieve the doubly-magic configuration near $^{132}$Sn.

A notable example outside the actinide region came in 2010, when experiments at the ISOLDE facility at CERN revealed that $^{180}$Hg, a nucleus for which the LDM predicts symmetric fission, undergoes distinctly asymmetric fission \citep{Andreyev2010}.
Unlike in the actinides, where the large spherical shell closures at $Z=50$ and $N=82$ in the fragments anchor the asymmetry, the $^{180}$Hg asymmetric split arises from comparatively small microscopic effects near the saddle point that render the fission valley mass-asymmetric, even though no strong fragment shell stabilization persists to scission.
That such subtle effects can decisively influence the mass split underscores the challenge of achieving predictive accuracy across the nuclear chart \citep{Andreyev2018}.

In addition to mass distributions, the charge distributions of fission fragments carry microscopic information.
A persistent odd-even staggering in the proton number yields, with even-$Z$ fragments produced more frequently than neighboring odd-$Z$ ones, provides evidence that nucleon pairing correlations survive the violent scission process \citep{Bocquet1989}.
The magnitude of this staggering varies systematically with the mass asymmetry of the split, offering further insight into how dissipation and pairing interact during the descent to scission \citep{Caamano2011}.

An important distinction in reporting fission yields is that between \emph{independent} and \emph{cumulative} yields.
Independent yields $Y(A,Z)$ give the production of a specific nuclide immediately after prompt neutron emission but before beta decay.
Cumulative yields additionally include contributions from all beta-decay precursors at the same mass number and are what most radiochemical measurements observe, since chemical separations typically occur on longer timescales.
Which quantity is relevant depends on the application: independent yields govern prompt radiation signatures and reactor kinetics, while cumulative yields determine long-term radiotoxicity and isotope production.

\subsection{Total kinetic energy}
\label{sec:tke}

The total kinetic energy (TKE) of the two fragments is the single largest component of the fission energy release, typically around 170~MeV for uranium fission.
Because the fragments acquire nearly all their kinetic energy from mutual Coulomb repulsion after scission, the TKE serves as a sensitive probe of the scission-point geometry.
A higher TKE implies a more compact configuration, with fragments closer together and a shorter neck, at the moment of rupture.

Empirical systematics, most notably those developed by \citet{Viola1985}, relate the average TKE to the charge and mass of the fissioning system:
\begin{equation}
  \langle\text{TKE}\rangle
\approx 0.1189\,\frac{Z^2}{A^{1/3}} + 7.3 \quad (\text{MeV}).
  \label{eq:viola}
\end{equation}
Deviations from this smooth trend reveal the influence of shell structure on the scission configuration.

The relationship between TKE and fragment mass is also revealing.
For $^{235}$U thermal fission, the TKE is highest (${\sim}185$~MeV) for mass splits near $A_H\approx 132$, where the heavy fragment is close to the doubly-magic $^{132}$Sn and adopts a compact spherical shape at scission, placing the two charge centers as close together as possible.
For more asymmetric or symmetric splits, the fragments are more deformed at scission, the charge centers are farther apart, and the TKE drops accordingly.
This correlation between TKE and fragment structure is a powerful diagnostic of the scission configuration.

The total excitation energy (TXE) of the system is the complement of TKE: $\text{TXE} = Q - \text{TKE}$.
The sharing of TXE between the two fragments is a question of active research.
Early models assumed equal temperatures, but recent experimental and theoretical work suggests that the sharing depends on the relative deformation of each fragment at scission and is therefore mass-asymmetric \citep{Talou2021}.

\subsection{Prompt neutron and gamma-ray emission}
\label{sec:prompt}

As described in Sec.~\ref{sec:scission}, the fragments emerge from scission hot and deformed.
Each fragment relaxes toward its equilibrium shape, converting deformation energy into additional internal heat, and then cools by evaporating prompt neutrons on a timescale of ${\sim}10^{-18}$--$10^{-14}$~s, followed by a cascade of gamma rays that carry off the remaining excitation energy and angular momentum.

\begin{figure}[htb]
\centering
\includegraphics[width=0.85\textwidth]{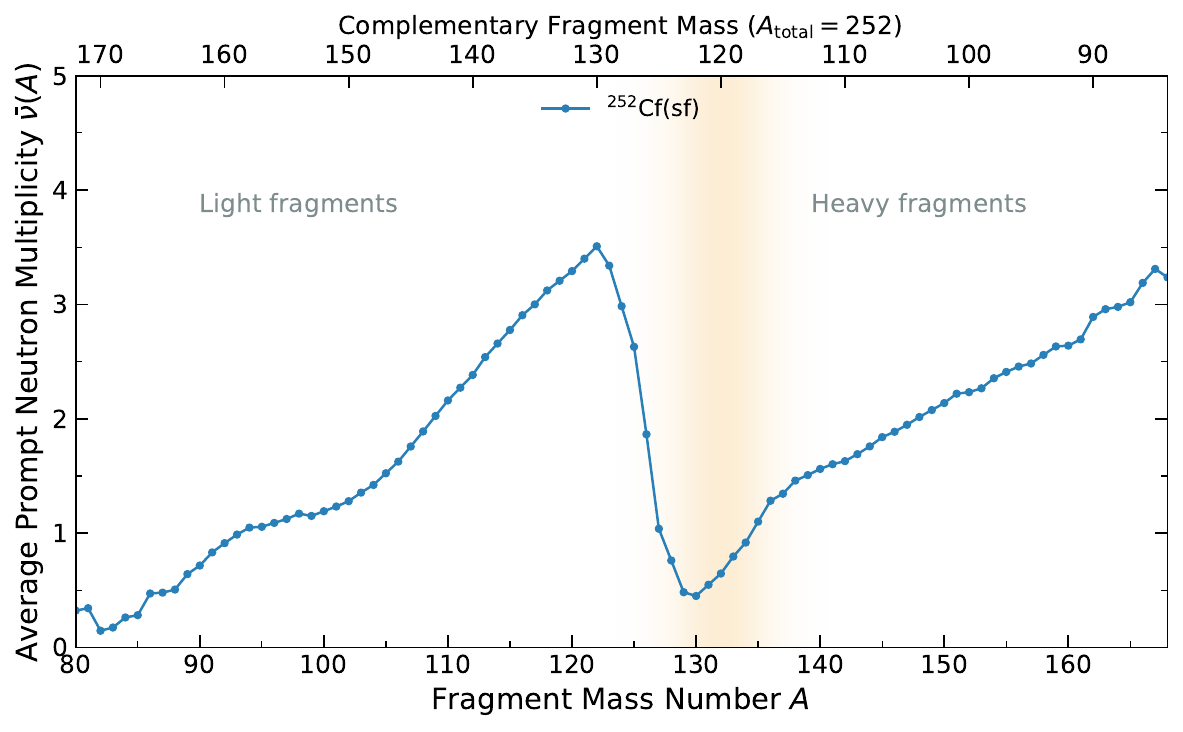}
\caption{Average prompt neutron multiplicity $\bar{\nu}(A)$ as a function
of fragment mass for $^{252}$Cf(sf), based on data from Ref.~\citep{BudtzJorgensen1988}.
The characteristic ``sawtooth'' shape reflects the role of shell structure: fragments near the doubly-magic $^{132}$Sn ($Z=50$, $N=82$) are nearly spherical at scission and emit very few neutrons (shaded region), while fragments farther from closed shells are more deformed and emit correspondingly more neutrons.}
\label{fig:sawtooth}
\end{figure}

The average number of prompt neutrons emitted as a function of fragment mass, $\bar{\nu}(A)$, displays the distinctive ``sawtooth'' pattern shown in Fig.~\ref{fig:sawtooth}.
Fragments near the doubly-magic $^{132}$Sn ($A\approx 130$--132) are nearly spherical at scission, so they acquire little deformation energy and evaporate very few neutrons ($\bar{\nu}\lesssim 0.5$).
Moving away from this shell closure in either direction, the fragments become increasingly deformed, absorb more energy upon shape relaxation, and emit correspondingly more neutrons.
The sawtooth shape thus provides a direct window into the deformation of fragments at scission and the role of shell structure in shaping fission outcomes \citep{Bowman1963}.

The prompt gamma-ray spectrum and multiplicity carry complementary information about the angular momentum of the primary fragments.
On average, each fragment emits ${\sim}7$--8 gamma rays with a total energy of ${\sim}7$~MeV.
The gamma-ray multiplicity is correlated with fragment mass in a pattern similar to (but smoother than) the neutron sawtooth \citep{Pleasonton1972}.

A key question is when and how the fragments acquire their angular momentum.
Recent measurements using the LICORNE directional neutron source coupled to the $\nu$-Ball gamma-ray spectrometer at the ALTO facility in Orsay demonstrated that the spins of the two partner fragments are uncorrelated and exhibit a strong mass-dependent sawtooth distribution, with no notable dependence on the identity of the partner nucleus \citep{Wilson2021}.
These findings indicate that fragment angular momentum is predominantly generated after scission, rather than being imparted by collective modes of the pre-scission nucleus.
The proposed mechanism involves the collective motion of nucleons in the ruptured neck region at the moment of scission, analogous to the snapping of an elastic band.
Theoretical work has since reproduced the sawtooth spin distributions and explored the underlying dynamics of angular momentum generation \citep{Randrup2021} the degree of spin correlation between partner fragments \citep{Randrup2026}, and the spatial orientation of fragment intrinsic spins \citep{Scamps2023}.

A particularly illuminating observable is the average total prompt energy release as a function of fragment mass.
By combining TKE, neutron, and gamma-ray measurements, one can reconstruct the complete energy balance of fission on a mass-by-mass basis.
The result confirms that the total $Q$-value is approximately constant for all mass splits of a given compound nucleus, but the partition between kinetic and excitation energy varies dramatically, a direct reflection of the different scission configurations explored by different fragment pairs.

\subsection{Delayed emission and beta-delayed neutrons}
\label{sec:delayed}

The primary fission fragments are generally neutron-rich and undergo a chain of beta decays toward stability.
In a small but crucial fraction of decays, the daughter nucleus is left with enough excitation energy to emit a neutron, known as a \emph{beta-delayed neutron}.
Although the delayed-neutron fraction $\beta_{\rm eff}$ is only about 0.65\% for $^{235}$U, these delayed neutrons are essential for reactor control.
To understand why, consider the neutron multiplication factor $k$, which must equal unity for steady-state operation.
If all neutrons were prompt (emitted within $10^{-14}$~s of scission), the reactor period, the timescale over which the neutron population changes, would be on the order of milliseconds, far too fast for any mechanical control system.
The delayed neutrons, emitted seconds after, slow the effective reactor period to a controllable timescale.
In practice, reactor design exploits this separation of timescales: the system is kept subcritical on prompt neutrons alone, so that criticality is achieved only when the delayed-neutron contribution is included.
This elegant safety feature exists because of the nuclear structure of the neutron-rich fission fragments and their beta-decay daughter nuclei, another example of fundamental nuclear physics underpinning a critical technology.
Each fissile isotope has its own characteristic delayed-neutron yield and time structure. 
Accurate knowledge of these parameters for all relevant fissile species ($^{235}$U, $^{239}$Pu, $^{233}$U, etc.) is a prerequisite for reactor kinetics calculations.

\section{Theoretical Frameworks}
\label{sec:theory}

Modeling fission from a theoretical standpoint is among the most demanding problems in nuclear physics.
It requires tracking the correlated motion of hundreds of strongly interacting fermions as they undergo a dramatic collective rearrangement.
Approaches range from phenomenological macroscopic--microscopic models, which have provided decades of quantitative predictions, to fully microscopic methods that aspire to describe fission from the underlying nuclear interaction.

\subsection{The macroscopic--microscopic method}
\label{sec:macmic}

The Strutinsky shell-correction approach described in Sec.~\ref{sec:shell} forms the backbone of the macroscopic--microscopic (mac-mic) method.
In practice, a smooth macroscopic energy functional (often the finite-range liquid drop model or FRLDM of M\"{o}ller \emph{et al.}\ \citep{Moller1995}) is augmented with shell and pairing corrections computed from single-particle levels in a deformed potential (typically a folded-Yukawa or Woods-Saxon form).
By evaluating this energy over millions of shapes spanning a high dimensional space of shapes, one constructs detailed potential energy surfaces, identifies saddle points and fission valleys, and extracts barrier heights.

The mac-mic method has been remarkably successful.
Large-scale calculations by M\"{o}ller, Sierk, and collaborators have mapped the fission barriers of nuclei across the entire chart of nuclides, over 8{,}000 nuclei from oxygen to the superheavy region, achieving agreement with measured barrier heights to within roughly 1~MeV \citep{Moller2009,Sierk1986}.
The FRDM (Finite-Range Droplet Model) mass table \citep{Moller1995}, which uses this framework, remains one of the most widely used nuclear mass models and serves as input to astrophysical simulations, reactor design codes, and nuclear data evaluations.

A particular strength of the mac-mic approach is its computational efficiency.
Because the macroscopic energy is given by an analytic formula and the shell correction requires only the diagonalization of a single-particle Hamiltonian (not a self-consistent iteration), PES calculations spanning millions of deformation-space grid points can be completed on modern workstations.
This speed allows systematic surveys over isotopic chains and the exploration of the high-dimensional deformation space needed to capture asymmetric fission pathways.

The principal limitation is that the mac-mic method is not fully self-consistent: the macroscopic energy and the single-particle potential are not derived from the same underlying interaction, and the potential parameters are fit to ground-state data.
Extrapolation to the extreme deformations and exotic neutron-to-proton ratios encountered in fission of neutron-rich nuclei carries unquantified systematic uncertainties.

\subsection{Nuclear density functional theory}
\label{sec:dft}

The microscopic approach to fission begins with nuclear density functional theory (DFT; for more details consult chapter on DFT), in which the many-body problem is recast in terms of local densities and currents rather than individual wave functions.
The most widely used implementation is the Hartree--Fock--Bogoliubov (HFB) method, which simultaneously accounts for the nuclear mean field and pairing correlations using phenomenological energy density functionals (EDFs) based on zero-range Skyrme or finite-range Gogny interactions \citep{Schunck2016}.

In the context of fission, constrained HFB calculations map out the PES by solving the self-consistent mean-field equations at each point in deformation space.
Because the EDF encodes the nuclear interaction in a form that is not tied to a particular mass region, DFT-based PES calculations can in principle be applied across the entire nuclear chart, including the neutron-rich nuclei needed for astrophysical applications.
In practice, however, current EDFs are calibrated primarily to ground-state properties of known nuclei, and the reliability of their extrapolations to exotic systems remains an open question; different functionals can yield substantially different barrier heights and fission pathways far from stability.
A practical limitation of static DFT is that the computational cost of self-consistent HFB solutions restricts most calculations to two-dimensional PES slices, whereas capturing the full complexity of fission requires two to five collective coordinates.
Moreover, as with the mac-mic approach, the PES alone provides only the energy landscape, not the dynamics; predicting fragment yields, kinetic energies, and excitation-energy sharing requires supplementing it with a dynamical framework.

It is worth noting the conceptual difference between the mac-mic approach and DFT.
In the mac-mic method, the macroscopic energy is an externally supplied functional (e.g., the liquid-drop or finite-range droplet model), and the shell correction is calculated separately from single-particle levels in a prescribed potential.
In DFT, both the bulk and shell contributions emerge self-consistently from a single energy density functional.
This self-consistency is a theoretical advantage: it ensures, for example, that the nuclear shape and density distribution are determined variationally rather than imposed by hand.
This self-consistency does not, however, guarantee superior extrapolation; the challenge of predicting fission far from the calibration data is shared by both frameworks \citep{Schunck2016}.

A recent development is the use of \emph{beyond-mean-field} techniques to incorporate quantum fluctuations.
The generator coordinate method (GCM) projects states of good particle number and angular momentum from the constrained HFB solutions and then superposes them to obtain a collective wave function that can tunnel through the barrier.
GCM calculations of fission-barrier heights and spontaneous-fission half-lives have shown promising agreement with data, though the computational cost increases rapidly with the number of collective coordinates \citep{Regnier2016}.
Recent work has extended these microscopic approaches to the dynamics of induced fission, providing a fully quantum mechanical description of the evolution from the compound nucleus to separated fragments \citep{Ren2022}.

\subsection{Dynamical approaches}
\label{sec:dynamics_theory}

\textbf{Time-dependent DFT (TDDFT)} One way to describe microscopic fission is time-dependent density functional theory, in which the full set of occupied single-particle wave functions is propagated in real time through the self-consistent mean field.
TDDFT makes no assumption about adiabaticity or predefined collective coordinates; the neck forms and ruptures as a natural outcome of the quantum evolution.
Pioneering calculations by Bulgac and collaborators have demonstrated that TDDFT can reproduce the gross features of fragment mass and kinetic energy distributions for $^{240}$Pu fission \citep{Bulgac2016,Bulgac2019}.

The power of TDDFT lies in its ability to capture phenomena that are inaccessible to adiabatic PES-based methods.
The neck rupture, for instance, emerges naturally as nucleon wave functions localize into one fragment or the other.
Pairing correlations, which make nuclear matter a superfluid, manifest as the entanglement of Cooper pairs across the neck just before scission, leading to characteristic odd-even effects in the fragment charge yields.
Fragment angular momentum, which determines the prompt gamma-ray emission, also emerges in TDDFT calculations, though the relative contributions of pre-scission collective modes and post-scission Coulomb excitation remain an active area of investigation (see Sec.~\ref{sec:prompt}).

However, each TDDFT trajectory represents a single stochastic realization of the fission event.
Generating statistically meaningful distributions of observables requires running hundreds to thousands of trajectories with different initial conditions, and the immense computational cost (millions of CPU-hours per event on leadership-class supercomputers) currently limits comprehensive surveys.
Ongoing algorithmic developments, including the use of wavelet bases, GPU acceleration, and machine-learning surrogate models, aim to bring the cost down by orders of magnitude.

\textbf{Langevin dynamics} A complementary and computationally more tractable approach treats the collective degrees of freedom (elongation, asymmetry, neck) as classical variables undergoing Brownian motion over the PES.
The Langevin equation,
\begin{equation}
  M_{ij}\,\ddot{q}_j
= -\frac{\partial V}{\partial q_i} - \gamma_{ij}\,\dot{q}_j + g_{ij}\,\xi_j(t)\,,
  \label{eq:langevin}
\end{equation}
couples the conservative driving force (gradient of the potential $V$) with a friction tensor $\gamma_{ij}$ and a stochastic force characterized by random variables $\xi_j(t)$ and strength $g_{ij}$, related to the friction and temperature via the fluctuation-dissipation theorem.
By running thousands of Langevin trajectories on a mac-mic or DFT-based PES, one obtains predicted distributions of fragment mass, TKE, and excitation energy.
The approach, developed extensively by Randrup, M\"{o}ller, and collaborators, has achieved impressive agreement with measured mass yields for a wide range of fissioning systems \citep{Randrup2011b,Sierk2017}.

The key inputs to a Langevin calculation are the PES (providing the conservative force), the inertia tensor $M_{ij}$ (quantifying the resistance of the nuclear shape to acceleration in each collective direction), and the friction tensor $\gamma_{ij}$.
The inertia is typically derived from microscopic quantum models of how individual nucleons behave as the nucleus deforms, while the friction is calculated from how these particles dissipate energy by interacting with the changing nuclear shape (recall the wall-window dissipation).
The stochastic force ensures that different trajectories sample a distribution of outcomes, naturally producing the broad mass and TKE distributions observed experimentally.
In the strongly overdamped limit, where friction dominates over inertia, the second-order Langevin equation reduces to a first-order random walk on the PES, termed \emph{Brownian shape motion} \citep{Randrup2011}.
This simplification eliminates the need for the uncertain inertia tensor entirely, requiring only the PES and a single global parameter (the critical neck size at which the mass split is determined) to produce remarkably accurate fragment mass distributions across a wide range of fissioning systems.

The Langevin framework also permits the study of multi-chance fission.
At high excitation energies, the compound nucleus may emit one or more neutrons before fissioning, each emission reducing the excitation energy and altering the effective PES.
These first-chance, second-chance, and higher-chance fission components contribute distinct mass distributions, and their relative weights depend sensitively on the competition between fission and neutron emission at each stage \citep{Mumpower2016beta,Mumpower2022}.

\subsection{Fragment de-excitation models}
\label{sec:deexcitation}

The primary fragments emerging at scission are characterized by their mass, charge, kinetic energy, excitation energy, spin, and parity.
Because fission fragments are typically very neutron-rich, they undergo a series of beta-minus ($\beta^-$) decays to reach the valley of stability.
Therefore, predicting the experimentally observed prompt neutrons and gamma rays requires simulating the statistical de-excitation of these fragments. 

Several Monte Carlo codes based on the Hauser--Feshbach statistical decay formalism \citep{Hauser1952} have been developed for this purpose.
The basic algorithm proceeds as follows: (1) sample the initial fragment mass, charge, TKE, excitation energy, and angular momentum from a distribution (either measured or computed from a scission-point or dynamical model); (2) at each step, compare the statistical decay widths for neutron emission, gamma-ray emission, and (where energetically possible) charged-particle emission; (3) select the decay channel stochastically according to these widths; and (4) repeat until the fragment has cooled below the neutron separation energy, after which it de-excites solely by gamma emission.

The three principal codes in current use are CGMF (Cascading Gamma-ray Multiplicity with Fission), developed at Los Alamos \citep{Talou2021}; FIFRELIN (FIssion FRagment Evaporation modeLINg), developed at CEA Cadarache \citep{Litaize2015}; and FREYA (Fission Reaction Event Yield Algorithm), developed at Lawrence Livermore \citep{Vogt2017}.
These codes differ in their treatments of the initial excitation-energy sharing between fragments, their nuclear level-density prescriptions, and their gamma-ray strength functions, but all produce event-by-event realizations of the full neutron and gamma-ray cascade.

\begin{figure}[htb]
\centering
\includegraphics[width=0.85\textwidth]{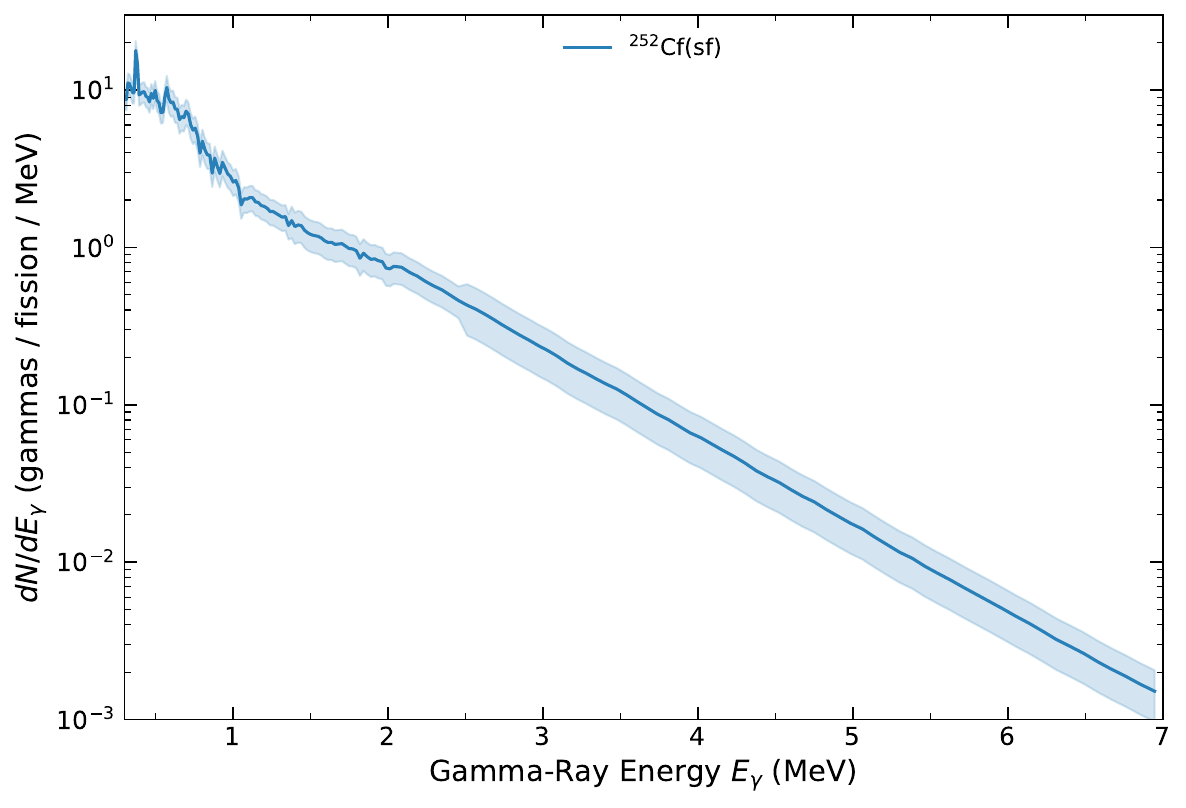}
\caption{Prompt fission gamma-ray spectrum $dN/dE_\gamma$ for $^{252}$Cf(sf), measured by \cite{Oberstedt2015}.
The shaded band indicates the experimental uncertainty.
The spectrum exhibits discrete transition structure at low energies ($E_\gamma \lesssim 1$~MeV) from collective rotational and vibrational de-excitation, transitioning to a smooth, approximately exponential continuum at higher energies dominated by statistical gamma emission.}
\label{fig:pfgs}
\end{figure}

The output of these de-excitation codes, including neutron multiplicity distributions, energy spectra, angular correlations, and gamma-ray spectra, can be directly compared with experimental data.
Figure~\ref{fig:pfgs} shows the measured prompt fission gamma-ray spectrum for $^{252}$Cf(sf), which spans nearly four orders of magnitude in intensity over the energy range 0.1--7~MeV and provides a stringent benchmark for these models.
They also serve as input to radiation-transport simulations used in reactor design, and nuclear security applications. 

As fission theory advances toward more first-principled scission-point distributions (from TDDFT or Langevin calculations), the coupling between dynamical fission models and de-excitation codes will finally provide a complete theoretical pipeline from the compound nucleus to the detected radiation.

\section{Experimental Methods}
\label{sec:experiment}

No single measurement provides all of the information needed to fully characterize fission: fragment masses, charges, kinetic energies, excitation energies, spins, and prompt emission must be pieced together from complementary techniques, each with its own strengths and blind spots.
Experimental fission physics has accordingly developed a diverse toolkit, which has undergone a transformation in the 21st century with new techniques providing unprecedented isotopic resolution and access to fission of short-lived nuclei far from stability.

\subsection{Traditional techniques}
\label{sec:traditional}

The earliest quantitative studies of fission relied on \emph{radiochemical} methods: irradiated targets were dissolved, chemically separated by element, and the resulting isotopes identified through their characteristic gamma-ray signatures.
Radiochemistry provided the first detailed maps of the double-humped mass yield curve and remains the basis of many entries in evaluated fission-product yield libraries.
Its strengths are excellent elemental ($Z$) identification and sensitivity to low-yield products; its limitations are that it is labor-intensive and cannot correlate the two partner fragments on an event-by-event basis.

The \emph{double-energy} ($2E$) method, developed in the early 1960s, became the workhorse of fission-fragment measurements \citep{Schmitt1966}.
A thin actinide target is placed between two back-to-back detectors, originally silicon surface-barrier detectors and later twin Frisch-grid ionization chambers \citep{BudtzJorgensen1988}.
Each detector records the kinetic energy of one fragment; conservation of momentum then yields the fragment masses.
The $2E$ method offers large solid-angle coverage and high count rates, but its mass resolution is limited by pulse-height defects in the detectors and energy-loss straggling in the target.

The \emph{double-velocity} ($2v$) method measures the time of flight of each fragment over a known path length using fast timing detectors.
Because prompt neutron evaporation is approximately isotropic in the fragment rest frame, it barely alters the fragment velocity, making the $2v$ method less sensitive to neutron-emission corrections than the $2E$ approach.
Early timing detectors, however, had limited resolution and small solid angles, which prevented the $2v$ method from supplanting the $2E$ technique as the standard.

More recently, the \emph{combined} $2E$--$2v$ technique measures both the kinetic energy and velocity of each fragment simultaneously \citep{AlAdili2018}.
The velocity measurement gives the pre-neutron-emission mass, while the energy measurement gives the post-neutron mass; the difference yields the prompt neutron multiplicity on an event-by-event basis without external input.
This concept was first realized by the COSI~FAN~TUTTE spectrometer at the Institut Laue-Langevin in the 1980s \citep{Boucheneb1989} and has been revived by modern instruments such as SPIDER, which offer substantially improved geometric efficiency \citep{Meierbachtol2015}.

All target-based methods share the limitation that the target must be fabricated from a sufficiently long-lived isotope, restricting measurements to a relatively small set of actinides.
The nuclear charge $Z$ of the heavy fragment is generally not measured directly, only inferred statistically.
Fission-fragment spectroscopy using large germanium detector arrays (such as GAMMASPHERE and EUROGAM) complements these kinematic methods by providing detailed rotational-band structures of individual fragments, giving access to their angular momenta, deformations, and single-particle configurations \citep{Hamilton1995}.

\subsection{Inverse kinematics}
\label{sec:inverse}

A significant experimental advance has been the use of inverse kinematics to study the fission process itself.
In this approach a heavy-ion beam is directed onto a light target, and because the fissioning system is moving at high speed, both fragments are boosted forward into a magnetic spectrometer that identifies their $Z$ and $A$ on an event-by-event basis.

Two complementary energy regimes are exploited.
At relativistic energies (${\sim}1$~GeV per nucleon), pioneered at GSI using the FRS separator and continued with the SOFIA experiment, electromagnetic excitation by a high-$Z$ target induces fission of the projectile \citep{Schmidt2000}.
At energies near the Coulomb barrier (${\sim}5$--10~MeV per nucleon), as implemented with the VAMOS++ spectrometer at GANIL, transfer and fusion reactions populate a variety of compound nuclei in a single experiment, providing access to multiple fissioning systems and excitation energies simultaneously \citep{Ramos2020}.
Both approaches achieve full isotopic identification of the fragments, impossible with traditional methods, and have opened a new window on the nuclear-structure information encoded in fission yields \citep{Andreyev2018}.

It is important to distinguish these dedicated fission studies from the use of in-flight fission as a \emph{beam production mechanism} at radioactive-ion-beam facilities such as FRIB in the United States and RIBF at RIKEN in Japan.
At these facilities, fission of relativistic uranium beams produces neutron-rich exotic isotopes that are separated and delivered as secondary beams for nuclear structure and reaction studies, but the fission process itself is not the primary object of investigation.

\subsection{Surrogate reactions and complementary techniques}
\label{sec:surrogate}

Many isotopes relevant to reactor design and astrophysics are too short-lived to be fabricated into targets.
The surrogate-reaction method circumvents this limitation by populating the same compound nucleus through an alternative reaction channel.
For example, neutron-induced fission of $^{237}$U (half-life 6.75~days) can be studied by the $(d,p\,f)$ reaction on a $^{238}$U target, which produces $^{239}$U$^{*}$ at a known excitation energy \citep{Escher2012}.

Storage rings at heavy-ion facilities provide another avenue, permitting precision mass measurements of fission isomers and short-lived fragments via their revolution frequencies.
At the Experimental Storage Ring (ESR) at GSI, for example, individual ions circulate for hours while their masses are determined from the revolution frequency to precisions of tens of keV, sufficient to resolve isomeric states and to provide stringent tests of mass models in the region of the chart most relevant to fission.

The CARIBU facility at Argonne National Laboratory takes a complementary approach: fission fragments from a $^{252}$Cf spontaneous-fission source are thermalized in a helium gas catcher and extracted as low-energy ion beams.
This chemistry-independent method delivers isotopically pure beams of neutron-rich nuclei for precision mass measurements, beta-decay spectroscopy, and post-accelerated reaction studies, providing nuclear-structure data in the heart of the fission-fragment region.
The recent nuCARIBU upgrade replaces the $^{252}$Cf source with neutron-induced fission of $^{235}$U, extending the range of accessible isotopes to different mass regions.

\subsection{Complementary techniques}
\label{sec:emerging}

Several additional techniques complement the methods described above.
Active-target time projection chambers (AT-TPCs), in which the detection gas also serves as the target material, allow the complete reconstruction of fission events including the emission angle and energy of all charged particles \citep{Ayyad2020}.
The SpecMAT detector at KU~Leuven and the ACTAR TPC at GANIL exemplify this approach.

Neutron time-of-flight facilities such as the LANSCE/WNR complex at Los Alamos, n\_TOF at CERN, and the Spallation Neutron Source at Oak Ridge provide neutron beams spanning energies from meV to hundreds of MeV, enabling precision measurements of neutron-induced fission cross sections over the full energy range relevant to both reactors and astrophysics.
Taken together, these complementary techniques collectively expand the reach of fission studies to the exotic nuclei and the extreme precision demanded by modern applications.

\section{Fission in the Cosmos}
\label{sec:astro}

Nuclear fission is not confined to terrestrial laboratories and reactors.
It has been proposed to play a pivotal role in the origin of the heaviest elements and their resultant electromagnetic signals in rare astrophysical events like neutron star mergers (see chapter on exotic explosions). 

\subsection{The $r$-process and fission recycling}
\label{sec:rprocess}

The rapid neutron-capture process ($r$-process) is responsible for producing a large share of the elements beyond iron, with current estimates attributing roughly half of the heavy-element inventory to this mechanism \citep{Burbidge1957,Cowan2021}.
In the $r$-process, seed nuclei capture neutrons far more rapidly than they can beta-decay, driving the nuclear composition along a path of very neutron-rich isotopes toward ever-heavier masses. (see chapter on nucleosynthesis processes) 
Because the vast majority of nuclei along the $r$-process path have never been produced in the laboratory, this process depends sensitively on theoretical predictions of nuclear masses, beta-decay rates, neutron-capture cross sections, and fission properties \citep{Mumpower2016}.

As nuclei climb to $A\gtrsim 260$, their fission barriers diminish and spontaneous, neutron-induced, and $\beta$-delayed fission can become the dominant mode of nuclear transmutations \citep{Giuliani2018}.
Fission thus acts as the ceiling of the $r$-process, preventing the formation of arbitrarily heavy nuclei \citep{Mumpower2018}.
The resulting fragments, with masses in the range $A\sim 100$--170, re-enter the neutron-capture chain as fresh seed nuclei.
This closed loop, termed \emph{fission cycling}, can operate multiple times during a single $r$-process event, efficiently redistributing nuclear material \citep{MartinezPinedo2007}.

The consequences of fission cycling for the final abundance pattern are profound.
The positions and shapes of the fission-fragment distributions determine which mass regions receive recycled material \citep{Beun2008}.
If the fragments are deposited predominantly near $A\sim 130$--140, they reinforce the second $r$-process peak and fill in the rare-earth region ($A\sim 160$).
The consistency of rare-earth abundances observed across metal-poor stars of widely different metallicities is considered circumstantial evidence for fission cycling \citep{Goriely2013}.

Careful analysis has demonstrated that the predicted $r$-process abundances in the $A\sim 100$--170 region depend on the fission-fragment distribution input \citep{Vassh2019}.
Different theoretical yield prescriptions, whether symmetric versus asymmetric or narrow versus broad distributions, produce qualitatively different abundance signatures, making $r$-process abundances a potential observational diagnostic of fission far from stability.

The specific fragment distributions produced by fission of neutron-rich superheavy nuclei, a regime beyond the reach of direct experiment, are among the most consequential unknowns in $r$-process nucleosynthesis \citep{Mumpower2020}.
Different theoretical fission-yield models can lead to qualitatively different predicted abundance patterns, motivating intense efforts to improve the predictive power of fission theory in this exotic territory \citep{Vassh2019}.

\subsection{Multi-messenger astronomy}
\label{sec:kilonova}

The detection of gravitational waves from a binary neutron-star merger in August 2017 (GW170817), together with the accompanying electromagnetic counterpart observed across the spectrum, provided the first observational hint that such mergers are major $r$-process sites \citep{Abbott2017}. Future correlated observations with neutrinos may help uncover a wealth of information associated with cataclysmic astrophysical events (see chapters on core collapse supernovae and supernova neutrinos,  supernova neutrino detection, stellar evolution and nucleosynthesis in stars, and neutron stars).

The optical and infrared transient that followed the merger, a \emph{kilonova}, was believed to be powered by the radioactive decay of freshly synthesized $r$-process nuclei in the expanding ejecta.
If actinides are created, at late times ($t\gtrsim 1$~day), the dominant heating source shifts from beta decay to spontaneous fission and alpha decay of heavy actinides and transuranic elements.
Fission deposits its ${\sim}200$~MeV of energy locally as kinetic energy of charged fragments, making it a particularly efficient heating agent \citep{Zhu2018}.

The opacity of the ejecta is also shaped by fission through the production of lanthanide and actinide elements, whose complex electronic structures give rise to a dense forest of absorption lines in the optical and near-infrared. Accurate modeling of light curves, and the emission of high-energy $\gamma$-rays therefore depends critically on fission-yield predictions for nuclei that exist only fleetingly in the most extreme environments in the universe \citep{Wang2020}.

While neutron-star mergers are currently considered the best candidate site for $r$-process nucleosynthesis, the question of whether supernovae also produce heavy $r$-process elements (and therefore involve fission cycling) remains open.
Magneto-rotationally driven supernovae and collapsars (rapidly rotating massive stars whose cores collapse to black holes) are candidate sites where sufficiently neutron-rich conditions might develop.
The role of fission in these environments depends on the electron fraction and expansion timescale of the ejecta, parameters that vary dramatically between models and require multi-dimensional simulations coupled with full nucleosynthesis networks that include fission \citep{Cowan2021}.

The connection between nuclear fission and astrophysical observables is mediated by many layers of physics: hydrodynamic expansion, radiative transfer, atomic opacities, and the full network of nuclear reactions all intervene between the fission event and the photons that reach a telescope.
Recent studies have shown that variations in nuclear physics inputs, including fission yields and barrier heights, can alter predicted light curves by orders of magnitude in brightness and shift their color evolution substantially \citep{Zhu2021,Barnes2021}.
Disentangling the fission contribution from these other uncertainties remains a major challenge, but multi-messenger observations of future neutron-star mergers offer a promising path toward constraining fission properties in a regime inaccessible to terrestrial experiments.

\section{Applications}
\label{sec:applications}

Beyond its role in fundamental research, fission underpins a range of technologies that directly impact society.

\subsection{Nuclear energy}
\label{sec:energy}

Commercial nuclear power, which supplies roughly 10\% of global electricity, exploits the self-sustaining chain reaction produced by fission.
In a thermal reactor, moderated neutrons induce fission primarily in $^{235}$U, and the resulting prompt and delayed neutrons sustain the chain.
Reactor operation is also affected by fission products that act as neutron poisons.
The most significant is $^{135}$Xe, which has the largest thermal-neutron absorption cross section of any known nuclide (${\sim}2.6\times 10^6$~b) and is produced copiously as a fission product.
Its buildup and decay following power changes must be carefully managed to avoid reactivity transients.

Current-generation light-water reactors (LWRs) operate with low-enriched uranium fuel (${\sim}3$--5\% $^{235}$U).
Next-generation designs, collectively termed Generation~IV systems, aim to improve sustainability, safety, and economics \citep{GenIV2014}.
Among these, fast-neutron reactors can exploit the \emph{breeding} cycle, in which $^{238}$U (which constitutes 99.3\% of natural uranium) captures a neutron and, following two beta decays, transmutes into fissile $^{239}$Pu:
\begin{equation}
  ^{238}\text{U} + n \to {^{239}\text{U}}
\xrightarrow{\beta^{-}} {^{239}\text{Np}} \xrightarrow{\beta^{-}} {^{239}\text{Pu}}\,.
  \label{eq:breeding}
\end{equation}
Once created, this newly bred $^{239}$Pu can absorb a subsequent neutron and undergo fission to generate power. 
By continuously converting the abundant but non-fissile $^{238}$U into combustible fuel, breeding effectively extends the uranium supply by nearly two orders of magnitude, from decades to millennia \citep{NEA2006}.

Advanced concepts also include small modular reactors (SMRs), which promise factory fabrication and deployment flexibility, and molten salt reactors (MSRs), which can operate with thorium fuel cycles.
In a thorium reactor, $^{232}$Th captures a neutron and transmutes to fissile $^{233}$U via two beta decays:
\begin{equation}
  ^{232}\text{Th} + n \to {^{233}\text{Th}}
\xrightarrow{\beta^{-}} {^{233}\text{Pa}} \xrightarrow{\beta^{-}} {^{233}\text{U}}\,.
  \label{eq:thorium}
\end{equation}
Similar to the plutonium cycle, this newly bred $^{233}$U then serves as the primary fissioning fuel for the reactor. 
Since thorium is roughly three times more abundant than uranium in the Earth's crust, this cycle could further extend the global nuclear fuel supply.
All of these advanced designs rely on detailed knowledge of fission observables, including cross sections, product yields, and delayed-neutron fractions, in non-standard fuel compositions and neutron energy spectra.
Uncertainties in the underlying nuclear data propagate directly into uncertainties in reactor performance and safety margins, driving a continued demand for both new measurements and improved theoretical models \citep{Kolos2022}.

\subsection{Medical isotope production}
\label{sec:medical}

Nuclear fission is the primary production route for $^{99}$Mo, the parent isotope of $^{99\mathrm{m}}$Tc (technetium-99m).
The daughter isotope $^{99\mathrm{m}}$Tc has a 6-hour half-life and emits a 140~keV gamma ray, properties that make it nearly optimal for single-photon emission computed tomography (SPECT).
It is the most widely used radioisotope in nuclear medicine, with global utilization estimated at over 30 million patient scans per year \citep{IAEA2017}.

Production typically involves irradiating highly enriched $^{235}$U targets in a research reactor, extracting the $^{99}$Mo from among the many fission products, and shipping it in the form of $^{99}$Mo/$^{99\mathrm{m}}$Tc generators to hospitals.
The fragility of this supply chain, which depends on a small number of aging research reactors, has prompted development of alternative production methods, including accelerator-based routes, but fission remains the workhorse for the foreseeable future.

\subsection{Transmutation of nuclear waste}
\label{sec:transmutation}

Spent nuclear fuel contains long-lived minor actinides, principally $^{237}$Np, $^{241}$Am, and $^{243}$Am, whose radiotoxicity dominates the long-term hazard of geological repositories.
Transmutation strategies aim to destroy these isotopes by fissioning them in fast reactors or accelerator-driven systems (ADS), converting them into shorter-lived or stable fission products. 
An ADS couples a high-power proton accelerator with a spallation neutron source and a subcritical core loaded with minor actinides.
The external neutron supply allows the system to operate safely below criticality while still achieving high transmutation rates.

The appeal of transmutation is compelling: the radiotoxicity of spent fuel, which in an open fuel cycle exceeds that of the original uranium ore for hundreds of thousands of years, can in principle be reduced to levels comparable to the ore within a few hundred years if the minor actinides are effectively fissioned.
However, realizing this technology requires accurate fission cross sections and yield data for minor actinides across a broad range of neutron energies, data that are often sparse or entirely absent.
The fission cross section of $^{241}$Am, for instance, is known to only ${\sim}5$--10\% precision above 1~MeV, an uncertainty that propagates directly into transmutation rate estimates and safety margins \citep{Salvatores2011}.
This underscores the continued need for both experimental measurements and improved theoretical predictions \citep{Forrest2011}.

\subsection{National security and nonproliferation}
\label{sec:security}

Nuclear fission has obvious connections to defense and nonproliferation.
The detection of fission signatures, including delayed gamma rays, antineutrinos from fission-product beta decays, and xenon isotope ratios, provides the technical basis for nuclear test monitoring, treaty verification, and reactor safeguards.
Accurate fission product yield databases are essential for forensic attribution of nuclear materials and for interpreting remote sensing data from monitoring stations.
Many of the same nuclear data needs that drive reactor design and astrophysics research also serve the nonproliferation community \citep{Romano2020}.

\section{Outlook and Open Questions}
\label{sec:outlook}

Despite more than eight decades of study, nuclear fission remains a field rich with open questions and emerging opportunities.

\textbf{Superheavy elements} The existence of elements beyond oganesson ($Z=118$) hinges on the competition between fission and alpha decay.
In the pure liquid-drop model, nuclei beyond $Z\approx 104$ should be instantly unstable against fission.
The fact that elements up to $Z=118$ have been synthesized is a testament to the stabilizing power of nuclear shell effects, which create a ``peninsula'' of relative stability extending from the known actinides into the superheavy region.
Predictions of a fully shell-stabilized ``island of stability'' near $Z=120$--126 and $N=184$ remain theoretically uncertain; different nuclear models disagree on the precise location and magnitude of the shell gap, and the fission barriers of these systems, calculated to be only a few MeV, are exquisitely sensitive to the details of the nuclear interaction.
Building on decades of superheavy-element synthesis at JINR Dubna \citep{Oganessian2015}, experiments at Lawrence Berkeley National Laboratory (LBNL) have recently demonstrated the production of livermorium ($Z=116$) using $^{50}$Ti beams \citep{Gates2024}, an essential step toward the synthesis of elements 119 and 120 that will directly probe the limits of nuclear existence imposed by fission.

\textbf{Microscopic predictive power} While TDDFT calculations have demonstrated proof of principle, achieving truly predictive, quantitative accuracy for fission-fragment distributions across the nuclear chart remains elusive.
Key challenges include the construction of improved energy density functionals, the treatment of quantum fluctuations and tunneling beyond the mean-field approximation, and the development of efficient algorithms for the massive computational demands of real-time many-body simulations \citep{Bulgac2020}.

\textbf{Fission of neutron-rich nuclei} The fission properties of nuclei along the $r$-process path, particularly at $A\sim 260$--300 and well beyond any experimentally accessible system, represent a critical gap in nuclear data.
For these nuclei, neither the barrier heights nor the fragment distributions are experimentally constrained; theoretical predictions differ by factors of two or more in key quantities such as the peak positions and widths of the mass yields.
Modern radioactive-ion-beam facilities such as FRIB in the United States, FAIR in Germany, and RIBF in Japan will extend measurements toward the neutron-rich frontier, but even the most exotic beams will fall short of the $r$-process path by many neutrons.
Large extrapolations will continue to rely on theory, and benchmarking theoretical models against new data in previously unexplored regions of the chart of nuclides will be essential for building confidence in those extrapolations.

\textbf{Connections to other fields} Fission dynamics shares deep mathematical structure with molecular dissociation: both involve a bound quantum system navigating a multidimensional potential energy surface to break apart into fragments, and both can be described by the Born-Oppenheimer framework and stochastic approaches such as Kramers' escape-over-a-barrier theory \citep{Kramers1940}.
Advances in these shared theoretical tools may provide cross-fertilization between nuclear fission theory and quantum chemistry.

\textbf{Precision for applications} The design and licensing of next generation reactors and transmutation systems demand fission data (cross sections, yields, prompt emission spectra) with uncertainties reduced to the percent level.
Quantified uncertainty propagation from nuclear data through simulations is an active and growing field \citep{Lovell2020}.

\textbf{Machine learning and data-driven methods} Emerging techniques in machine learning are being applied to interpolate and extrapolate fission observables, to emulate expensive DFT calculations, and to optimize experimental campaigns \citep{Boehnlein2022}.
Neural-network-based emulators are being built that can approximate the output of a full HFB calculation in milliseconds, making it feasible to explore parameter spaces that would otherwise require years of supercomputer time.
Bayesian optimization methods are being used to design maximally informative experiments \citep{Siefman2021}. 
By selecting the target nuclei, beam energies, and detector configurations that will most efficiently reduce uncertainties. 
While still in early stages, these data-driven methods hold promise for accelerating the convergence of both theory and experiment toward the predictive accuracy demanded by applications.

\textbf{Nuclear data evaluation and uncertainty quantification} The bridge between fundamental fission research and practical applications passes through evaluated nuclear data libraries such as ENDF/B (United States), JEFF (Europe), and JENDL (Japan).
These libraries compile, evaluate, and recommend ``best values'' for fission cross sections, product yields, and prompt emission data based on all available experimental and theoretical information.
A growing emphasis in the evaluation community is the rigorous quantification of uncertainties and their covariances, essential for propagating nuclear data uncertainties through reactor simulations and assessing safety margins.
The interplay between new measurements, improved theory, and modern evaluation methodology defines the path toward the precision fission data needed for the next generation of nuclear technologies.

Nuclear fission continues to challenge and inspire.
It connects the quantum mechanics of strongly interacting fermions to the macroscopic energy output of power reactors and to the elemental composition of the cosmos.
As experimental facilities push toward more exotic nuclei and theoretical methods grow in sophistication, the field is poised for discoveries that will deepen our understanding of the atomic nucleus and expand the applications that depend on it.


\bibliographystyle{Harvard}
\bibliography{references}

\end{document}